\documentclass[preprint,aps]{revtex4-1}
\usepackage{amsmath}
\usepackage{url}
\usepackage{amsfonts}
\usepackage{amssymb}
\usepackage{textcase}
\usepackage{bm}
\usepackage{braket}
\usepackage{color}
\usepackage{graphicx}
\begin{document}
\title{Cloaked Resonant States in Bilayer Graphene}
\author{A. V. Shytov}
\affiliation{
School of Physics, University of Exeter, Stoker Rd, Exeter, EX4 4QL, 
United Kingdom
}
\begin{abstract}
Charge carriers in bilayer graphene  occupy two parabolic 
continua of electron-like and hole-like states which differ by 
the alignment between carrier pseudospin and its momentum, 
the property known as chirality. Due to chirality conservation, 
a strong confining potential can  host unusual bound states: 
electron levels cloaked into the hole continuum. The energy levels and 
the wave functions of  the cloaked states can be obtained 
by solving the Schr\"odinger equation for a massive non-chiral 
particle in the~$p$-wave channel in two dimensions.  
Eventually, cloaked states slowly decay into the continuum, 
via  trigonal warping effects. 
We discuss the key properties of cloaked states in circularly 
symmetric potentials, and show that cloaking should be observable  
in quantum corral  geometries via scanning tunneling probe measurements. 
\end{abstract}
\maketitle

There are two related fundamental facts that make the electronic properties of 
graphene interesting: the coexistence of electron and hole states
in the vicinity of conic points in the spectrum, and chirality of 
charge carriers. Charge carriers in graphene can be characterized
by their pseudospin, and different alignments between pseudospin
and momentum result in electron-like and hole-like states. 
There is an important difference between the cases of monolayer and
bilayer graphene (BLG) due to their different dispersion laws
and different structure of chiral states. 
In single-layer graphene, the linear dispersion results in the 
density of states vanishing at the Dirac point~\cite{MLG-dos-properties}.  
In bilayer graphene, 
the dispersion is parabolic, and the density of states is
flat~\cite{Falko}, so that
there is a large number of both electron- and hole-like states near
the point separating the two subbands. One consequence of this 
is a potential instability of bilayer graphene with respect to formation
of broken-symmetry phases~\cite{Broken-Phases}. 
The chirality also affects transport properties. In monolayer graphene, 
any potential barrier is perfectly transparent at normal
incidence~\cite{Klein-paradox}. In bilayer graphene, chirality is 
responsible for electron cloaking~\cite{LL-Cloaking} described below. 

The effects of chirality are manifested in the properties of bound 
and quasi-bound states. In monolayer graphene, 
a strong confining potential, such as the supercritical Coulomb
potential\cite{Coulomb-Atomic-Collapse}, can give rise to a quasi-localized
resonance which was recently observed in scanning tunneling microscopy
(STM) experiments\cite{Crommie-Supercritical}. The resonant state is 
formed by electron-like states submerged into  the hole band
and coupled to it via Klein tunneling. 
The number of states available for decay is suppressed by linear
density of states, so that the width  of the resonant peak  
is smaller than its energy, and the resonance 
can be resolved. 

Do  localized resonances exist in BLG? At first glance,  
the large flat density of states in BLG would result in the fast 
decay of such a state. However, the recent analysis of 
of barrier transmission problem in bilayer graphene~\cite{LL-Cloaking} 
suggests that the coupling to the continuum can be suppressed 
as a consequence of chirality conservation. 
For a one-dimensional barrier~$U(x)$, 
the quasimomentum perpendicular to the barrier, $p_y$, is conserved. For
normal incidence, $p_y = 0$, electron and hole continua decouple, and the 
problem is equivalent to two one-dimensional Schr\"odinger equations
for two massive, non-chiral particles, which differ only by
the sign of the potential energy. An attractive potential~$U(x) <
0$ hosts at least one bound state at negative energies, for any
potential strength. The
state is submerged into the continuum of holes, repelled 
by the potential~$-U(x)$. In~\cite{LL-Cloaking} this phenomenon was
referred to as {\it cloaking}, by analogy with cloaking in
optics~\cite{Cloaking-Pendry, Cloaking-Leonhardt}.  
Indeed, 
one cannot probe, e.g., the occupancy of  the level 
using decoupled hole states coexisting in  energy. 
Away from normal incidence, however, the coupling 
between electron and hole states is proportional to~$p_y$, so that
the localized level acquires finite  width~$\propto
p_y^2$, and thus is turned into a resonant barrier state
which can contribute to the transmission~\cite{Campos-RBS}. 
Thus, a cloaked state appears as a singular point at~$p_y = 0$.
Another manifestation of decoupling between electrons and holes in 
quantum dots in graphene was mentioned in the context of 
quantum dots in BLG~\cite{Peeters}.

In this Letter, we show that cloaked states in BLG can 
be hosted by a local, circularly symmetric potential.
The cloaking occurs in the $s$-wave 
channel, which serves as an analogue of the normal incidence channel
in the barrier transmission problem. Thus, the cloaking relies on 
a discrete quantum number, which results in perfect isolation of the
cloaked state from the continuum, and gives a truly discrete level. 
Unlike the case of one dimension, however, we also find that the 
bound states only occur if the potential exceeds a certain critical 
strength. We also consider how the cloaked states decay due to 
trigonal warping, and show that their  resonant 
width is less than~$1$meV. We also discuss potential experimental 
signatures of cloaked states. 

To analyze the dynamics of charge carriers in BLG, we 
introduce the following single-particle Hamiltonian\cite{Falko}
\begin{equation}
{\hat H} = \frac{1}{2m^\ast} 
\left(
\begin{array}{cc}
0 & ({\hat p}_x - i {\hat p}_y)^2 \\
({\hat p}_x + i {\hat p}_y)^2 & 0 
\end{array}
\right)
 + U (x, y) 
\ ,
\label{bi-h}
\end{equation}
where~${\hat p}_{x, y} = - i \hbar \partial_{x, y}$ are the standard 
momentum operators, $m^\ast \approx 0.036m_e$ is the effective mass
of charge carriers, 
and~$U(x, y)$
is the confining potential. 
The matrix Hamiltonian operates on two-component wave functions 
which represent the two pseudospin states. 
For free electrons ($U = 0$), the eigenstates
of this Hamiltonian give two parabolic dispersion bands,
$\epsilon_{\bf k} = \pm {\hbar^2 {\bf k}^2}/({2 m^\ast})$. 
The chirality of charge carriers is reflected in the non-trivial phase 
of the momentum combination~$({\hat p}_x + i {\hat p}_y)^2$. 
In particular, this
leads to the relative momentum-dependent phase  between the two 
components of a plane wave: $\Psi_2 = \pm e^{- 2i \phi_{\bf k}}\Psi_1$, 
where~$\phi_{\bf k}$ is the azimuthal angle of the wavevector~${\bf k}$.

We analyze the case of a circularly symmetric potential, 
introducing the polar coordinates: $x = r \cos\phi$, $y = r\sin\phi$, 
so that~$U = U(r)$. 
The quantum states can be characterized by their angular momentum quantum
number~$M = {\ldots}, -2, -1, 0, 1, 2, {\ldots}$, which takes integer values.
We shall represent the wave function as a superposition of two components 
of opposite chirality described by the two amplitudes, 
$u_M(r)$ and~$v_M(r)$: 
\begin{equation}
\Psi_{M}(r, \phi) = e^{i M \phi} 
\left(
\begin{array}{r}
e^{-i \phi}  [u_M(r) + v_M(r)]  \\
e^{i\phi}  [u_M(r) - v_M(r)]
\end{array}
\right)
\ .
\label{psi-ansatz}
\end{equation} 
Note that the magnetic number for the two pseudospin components
differ by two; this is characteristic of bilayer graphene. 
Let us substitute the ansatz~(\ref{psi-ansatz}) into Schr\"odinger
equation~$\hat{H}\Psi_M = \epsilon \Psi_M$ 
with the Hamiltonian~(\ref{bi-h}). In polar coordinates, 
the differential operators
$\partial_x \pm i \partial_y$ take the form~$e^{i \phi} \left[
\partial_r \pm \frac{i}{r} \partial_\phi\right]$. 
%A trivial calculation\footnote{for ref. purposes, move to
%appendix/supplement/etc} yields
%\begin{eqnarray}
%(\partial_x + i \partial_y)^2 e^{i (l - 1) \phi} r (u_l(r) + v_l(r)) 
%= e^{i (l + 1) \phi} r \left(\partial_r^2 + \frac{1 - 2l}{r}
%\partial_r + \frac{l^2 - 1}{r^2}\right) (u_l  + v_l) 
%\ ,
%\\
%(\partial_x - i \partial_y)^2 e^{i (l + 1) \phi} r (u_l(r) - v_l(r)) 
%= e^{i (l - 1) \phi} r \left(\partial_r^2 + \frac{1  + 2l}{r}
%\partial_r + \frac{l^2 - 1}{r^2}\right) (u_l  - v_l) 
%\ .
%\nonumber
%\end{eqnarray}
The chiral components~$u_M(r)$ and~$v_M(r)$ can then be separated by 
calculating the sum and difference of the two equations:
\begin{eqnarray}
\label{uv-eqns}
\left[\epsilon - U\right] u_M  &=& - \frac{\hbar^2}{2 m^\ast}
\left[
{\hat {\cal D}}_M u_M 
- \frac{2M v_M'}{r} \right]
\ , 
\\
- \left[\epsilon - U\right] v_M  &=&   - \frac{\hbar^2}{2 m^\ast}
\left[
{\hat{\cal D}}_M v_M 
- \frac{2M u_M'}{r} \right]
\ ,
\nonumber
\\
{\rm where}
\quad
{\hat {\cal D}}_M u &\equiv& u'' + \frac{u'}{r} + \frac{M^2 - 1}{r^2} u
\ , u' \equiv \frac{du}{dr}
\ .
\nonumber
\end{eqnarray}
The terms proportional to~$2M/r$ describe the coupling between~$u_M(r)$
and~$v_M(r)$ and makes this system equivalent to a biharmonic
equation. 
Interestingly, the equations decouple in the $s$-wave channel, $M = 0$:
\begin{eqnarray}
\label{uv-decoupled}
\epsilon u_0  &=& - \frac{\hbar^2}{2 m^\ast} {\hat {\cal D}}_0 u_0
+ U(r) u_0 
\ , 
\\
- \epsilon v_0  &=&  -  \frac{\hbar^2}{2 m^\ast}
{\hat {\cal D}}_0 v_0
- U(r) v_0 
\ .
\nonumber
\end{eqnarray}
These equations resemble the ones considered in~\cite{LL-Cloaking}, 
in the limit of normal incidence. 
The two chirality components
are described by the two decoupled amplitudes~$u_0(r)$ and~$v_0(r)$; 
the confining  potential~$U(r)$ is attractive for one component 
and is repulsive for the other. 
The differential operator~${\hat {\cal D}}_0$ on the right-hand side of
Eqs.(\ref{uv-decoupled}) 
is identical to  the radial part of the Laplacian operator in two dimensions 
in the angular momentum channel with~$M = \pm 1$, i.e., in the~$p$-channel. 
Therefore, the solutions in the $s$-wave channel for bilayer graphene 
could be constructed from $p$-wave solutions of the Schr\"odinger equation
for a massive non-chiral particle in two dimensions. (Alternatively, 
one can map the wave function onto the~$s$-state in four dimensions, by the
change of variable~$u_0(r) = r w_0(r)$, which brings the kinetic
energy to the form~$w_0'' + 3 w_0'/r$.) 
For positive chirality states 
described by~$u_0(r)$
the potential in the Schr\"odinger equation is~$U(r)$, and the energy
is~$\epsilon$; for states in the opposite chirality channel 
these quantities change
sign. Thus, the two components, $u_0(r)$ and~$v_0(r)$ describe the 
two $s$-wave subbands of electron-like and hole-like states, respectively.

Before  employing this mapping, let us recall the  properties of bound states
for a massive non-chiral particle.  In dimensions less than or 
equal to two, a bound state in the~$s$-channel 
exists for an arbitrary weak potential; however, the binding energy
decreases with increasing dimension~\cite{LL-low-d}.
Bound states in higher
momentum channels exist only if the potential is sufficiently strong, 
so that it  can overcome quantum zero-point motion energy and the 
centrifugal barrier. 
The massive particle in more than two dimensions can be confined by a
negative potential~$U(r)$ if its strength exceeds a certain critical value, 
of the order of the zero-point energy in the potential.
For a three-dimensional square well~\cite{LL-bound} of 
radius~$a$ ($U(r) = -U_0$ for $r < a$), this value
is~$U_{0, {\rm cr}} = \frac{\pi^2 \hbar^2}{8 m^\ast a^2}$.
The same analysis can be carried over for a bound state in 
the~$p$-channel in two dimensions: the bound 
state becomes detached from  the continuum 
when the wave function inside the well 
reaches its maximum at the  well boundary. 
On the other hand, the wave function in the $p$-channel regular at~$r
= 0$ must be proportional to~$r$. 
For a rectangular well, the wave function inside the well 
is~$J_1(q r)$, where~$q = \sqrt{2 m^\ast U_0}/\hbar$ 
is the wavenumber inside the well, 
and~$J_1(x)$  is the first order Bessel function. 
This function reaches its first maximum at~$x = 1.84$, which
gives the critical well depth~$U_{0, {\rm cr}} = 1.70 \hbar^2 /
(m^\ast a^2)$. The estimate~$U_{0, {\rm cr}} \sim \hbar^2 /
(m^\ast a^2)$ is valid irrespectively of the details of the well
profile.

We can now state the central finding of this paper: discrete 
bound states exist in the electron spectrum of bilayer
graphene when the  potential~$U(r)$ exceeds the critical
strength: if it hosts a bound state 
in the $p$-channel for  the massive non-chiral particle
in two dimensions, it can also host a cloaked bound state in BLG. 
The critical potential strength is of the order of the kinetic 
energy of quantum zero-point motion, $\hbar^2/(m^\ast
a^2)$, where~$a$ is the radius of the potential.  
Note, however, that the bound state in bilayer graphene occurs 
irrespective of  the sign of the potential energy, due to the
particle-hole symmetry. For an attractive 
potential, 
the cloaked state is split off the electron continuum and submerged 
into the hole continuum, while a repulsive potential gives rise 
to a hole-like state. 

\begin{figure}
\includegraphics[width=0.45\textwidth]{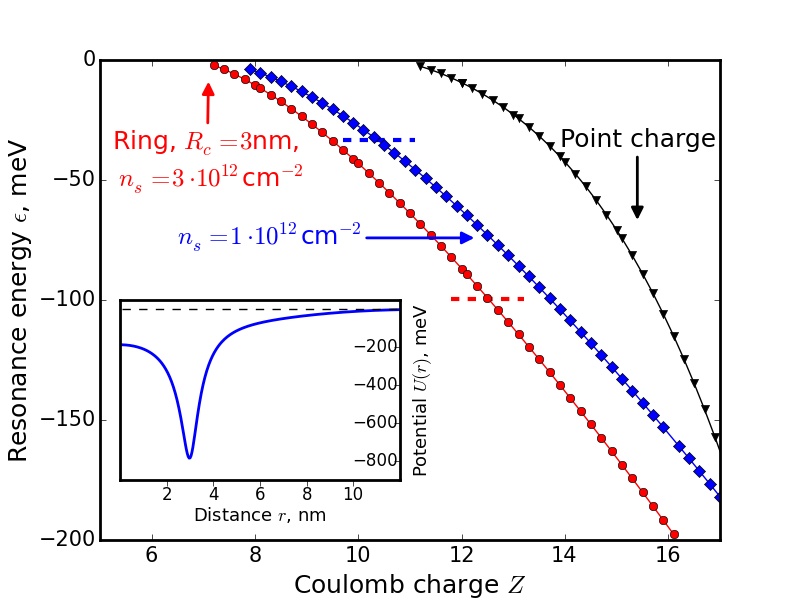}
\caption{
\label{fig-Z-dep}
The position  of the cloaked level as a function of the total charge for
the two distributions: (i) point-like charge (black), (ii) charged ring
(red, blue).  One can see that the cloaking is facilitated by
deconcentrating the charge.  The screening is slightly affected by 
the background charge density~$n_s$: for the lower charge density (the blue
curve), the screening is enhanced at short distances owing to the 
wavevector dependence of the polarization operator~\cite{Das-Sarma-Pi}. 
This results in smaller energy of the cloaked state. 
The dashed lines indicate the position of the Fermi level. 
Inset: the potential energy profile~$U(r)$ in the ring configuration
for the background charge density~$n_s = 3 \cdot 10^{12}$cm${}^{-2}$.  
}
\end{figure}

Let us now discuss the physical situations in which the cloaked states
can occur. It is very natural to check if cloaked states can be hosted
by charged impurities. One can consider the cloaking by the Coulomb
potential~$U(r) = -Ze^2/\varkappa r$, induced by the charge~$Ze$, 
where~$\varkappa$ is the effective dielectric constant of graphene and
its environment. 
For such a potential, infinitely many Rydberg-like bound  states 
exist at arbitrarily 
small charge~$Z$ in all momentum channels. In principle, a Coulomb 
potential acting on electrons in bilayer graphene would also host 
a hydrogen-like spectrum of cloaked states~\cite{hydrogen-2d}, with 
energies~$\epsilon_n = 
- E_R Z^2 / (n + 3/2)^2$, with Rydberg energy~$E_R = \frac{1}{2}  m^\ast e^4 / 
(\varkappa \hbar)^2$.   
For bilayer graphene on a SiO${}_2$ substrate ($\varkappa = 2.5$), 
the energy of the lowest cloaked
state would be~$\epsilon_0 =  -35$meV, 
due to small effective  mass. 
However, unlike the case of donor states in semiconductors, an external Coulomb
potential is effectively screened in bilayer graphene due to the non-vanishing
density of states and~$N = 4$ fermion flavours. The corresponding 
Thomas-Fermi screening 
radius, $\hbar^2 \varkappa / (N m^\ast e^2) \approx 0.37$nm 
is therefore shorter than  
the Bohr radius~$\hbar^2 \varkappa / (m^\ast e^2) \approx 1.5$nm.
The screening by charge carriers in BLG is mostly similar to 
Thomas-Fermi screening~\cite{Das-Sarma-Pi}, except  at distances shorter than 
the Fermi wavelength,  where the response is larger by 
the factor of~$\log 4 \approx 1.4$.
This response function can be employed to find the potential induced by 
a point charge, and  then solve Eq.~(\ref{uv-decoupled}) 
for~$u_0(r)$ numerically for different values of the Coulomb charge~$Z$. 
The resulting dependence of the cloaked level position on~$Z$ 
is shown as the black curve in  Fig.~\ref{fig-Z-dep}. 
The cloaked state exists if~$Z$  exceeds the critical value~$Z_c \approx 12$.
Therefore, it is very  unlikely that cloaked  states arise as donor 
or acceptor states  due to charged impurities. The culprit is 
the repulsion by a centrifugal barrier in the~$p$-state, 
which prevents the particle from reaching the attractive core
of the potential. 

\begin{figure}
\includegraphics[width=0.48\textwidth]{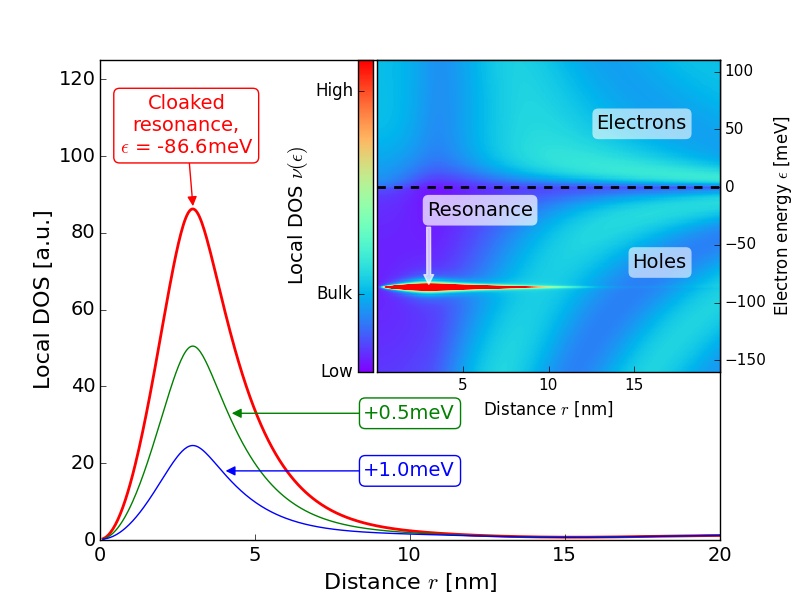}
\caption{
\label{fig-ldos}
The Local density of electronic states (LDOS) 
in quantum corral formed by a ring
of charged impurities ($3$nm in radius, total charge~$Z = 12$, 
background electron density~$n_s = 3 \cdot 10^{12}$cm${}^{-2}$) 
was obtained by solving Eq.(\ref{uv-eqns}) numerically for all 
channels with~$|M| < 10$. The trigonal warping contributions were
included  for the~$M = 0, 3, -3$ channels.  
The LDOS is shown in the units of  
its bulk value, $\nu_0 = m^\ast / (2\pi \hbar^2)$. 
Near the resonant energy, the LDOS 
exhibits a sharp peak at~$\epsilon = -86$meV 
that represents the probability distribution 
in the cloaked state. The peak dominates over the surrounding
continuum due to its small width, $0.64$meV.
Inset: the LDOS as a function of energy and the distance away from 
the center of the corral. 
The resonant peak is seen as a sharp streak superimposed onto the smooth
interference fringes formed by incident and repelled waves 
in the hole continuum. 
}
\end{figure}

The estimate for the critical potential strength, $U_{\rm cr} \sim
\hbar^2/(m^\ast a^2)$, shows that cloaked states can arise in 
smooth potentials with a large characteristic distance~$a$.
For example, if~$a = 10$nm, the critical strength is estimated
to be of the order of~$50$meV. 
One can also notice that distributing the positive charge 
rather than concentrating it at a single point is a way to 
get around the centrifugal barrier. This suggests a better
way to engineer cloaked states. One can create a large cluster
of charge impurities by manipulating them with the tip of a scanning
microscope. Arranging the impurities into a charged
ring (such as a quantum corral~\cite{Quantum-Corrals}), one can lower 
the required threshold.  For example, for impurities placed around
the ring of~$3$nm radius, the simulations give a lower value of the
critical charge, ~$Z_c = 7$, weakly dependent on the background
electron density, as shown by the red and blue curves in
Fig.~\ref{fig-Z-dep}. 
It should be also possible to engineer cloaked states via gating a BLG 
sample locally, e.g, with the tip of a scanning tunneling 
microscope~\cite{Whisper-gallery}. 

To detect the cloaked resonance, one can measure the local density of
states (LDOS) via scanning tunneling microscopy measurements.
The LDOS can be defined in terms of the wave
functions~$(u_{M, \epsilon}(r), v_{M, \epsilon}(r))$ 
at a given energy~$\epsilon$ as $\nu(\epsilon, r) 
\propto sum_{M} [u_{M, \epsilon}^2(r) + v_{M, \epsilon}^2(r)]$. 
To show how the bound state could reveal itself in STM measurements, 
we have calculated the LDOS numerically, including the effects
of  RPA screening. The decay due to trigonal warping (see below) was included
as Lorentzian broadening of the level. 
Figure~\ref{fig-ldos} shows the LDOS in the vicinity of 
a charged ring with total charge~$Z = 12$. 
The resonant contribution can be seen as a sharp fin at~$\epsilon_c \approx
-86$meV sticking out of the hole continuum.

We now turn to the issue of stability of cloaked states.
Deviations of the kinetic energy from the chiral form~(\ref{bi-h}) 
results in the decay of the bound states. The most important
next-order
contribution to the kinetic energy is the trigonal warping
term~\cite{Falko}: 
\begin{equation}
H_{w} = v_3 \left( 
\begin{array}{cc}
0 & {\hat p}_x + i {\hat p_y} \\
{\hat p}_x - i {\hat p_y} & 0 
\end{array}
\right)
\ , 
\end{equation}
where the parameter~$v_3 \approx 6 \cdot 10^4$m/s. 
The warping breaks chirality and angular momentum conservation.  
Applying~$H_w$ to the wave function~(\ref{psi-ansatz}), 
we find
\begin{equation}
H_{w} \left(\begin{array}{c}
e^{-i\phi}  \\
e^{i\phi}
\end{array}
\right) u_0(r) = -i \hbar v_3 \left[u_0' - \frac{u_0}{r} \right]  \left(
\begin{array}{c} 
e^{2 i \phi} \\
e^{-2i \phi} 
\end{array}
\right)
\ .
\end{equation}
Hence, the cloaked~$M = 0$ state could decay into~$M = \pm 3$ continuum
states. This decay, however, is partially  suppressed by two effects. 
Firstly, 
the wave function with large angular momentum is suppressed near the
origin~$r = 0$ due to the centrifugal barrier. Second, the continuum
state, unlike the cloaked state, is repelled by the
potential~$-U(r)$. Near the critical threshold, 
when the resonant state is shallow, the dominant effect 
is due to the centrifugal potential. 
A simple estimate of the resulting resonant width can be found with the 
help of Fermi's golden rule.
Let us assume, for simplicity, that we consider the state near the 
cloaking threshold. This state can be described by a shallow 
wave function, independent of the profile of the potential. 
For the~$p$-channel, the decaying shallow-state wave function 
is proportional to the first-order Macdonald function~$K_1(\kappa r)$. 
The parameter~$\kappa$ is related to the energy
of the cloaked state: $\epsilon_c = - \frac{\hbar^2 \kappa^2}{2 m^\ast}$
The outgoing wave that coexists at the same energy has the 
wavenumber~$\kappa$, and the corresponding wave function in the~$M = 3$ 
channel can be written in terms  of Bessel functions~$J_{M \pm
1}(\kappa r)$. The properly normalized initial and final states are
\begin{eqnarray}
\Psi_{\rm in} 
&=& \frac{\kappa K_1(\kappa r)}{\sqrt{4\pi \log \frac{1}{\kappa a}}} 
\left(
\begin{array}{r} 
e^{-i\phi} 
\\ 
e^{i\phi} 
\end{array}
\right)
\ , 
\\
\nonumber
\Psi_{\rm out} &=& 
\sqrt{\frac{\kappa}{4 R}}
\left(
\begin{array}{c}
J_2 (\kappa r) e^{2i \phi} \\
J_{4} (\kappa r) e^{4 i \phi}
\end{array}
\right)
\ , 
\end{eqnarray}
where~$R$ is a large normalization radius for continuum states. 
Applying recursive relations between Bessel functions 
and known integrals~\cite{Gradstein},  one can find the transition 
matrix element: 
\begin{equation}
V \equiv \left\langle \Psi_{\rm out} | {\hat H}_w | \Psi_{\rm in} 
\right \rangle 
= -i \hbar v_3 \sqrt{\frac{\pi \kappa}{16 R \log\frac{1}{\kappa a}}} \ .
\end{equation}
The density of final states in a given angular momentum channel is
~$\nu_{M = \pm 3} 
= (R/\pi) ({d\kappa}/{dE}) = (R m^\ast)/({\pi \hbar^2
\kappa})$. The golden rule formula then gives the decay width as 
\begin{equation}
\Gamma = 2 \times \frac{2\pi}{\hbar} |V|^2 \nu_{M = 3} = \frac{\pi}{4
\log\frac{1}{\kappa a}} m^\ast v_3^2
\end{equation}
(we have included both momentum channels here). 
Thus, we see that the broadening of the cloaked state
is only weakly dependent of its energy, and is given by the 
quantity~$(\pi/4) m^\ast v_3^2 \approx 0.6$meV. 
Therefore, the warping becomes  insignificant when the energy
of the cloaked state exceeds this value. 
By dimensional arguments, this estimate should hold, by the order of
magnitude, irrespective of the details of the well~$U(r)$. 

One should also bear in mind that transverse electric fields 
can open a gap in bilayer graphene~\cite{Bilayer-Gap}.
The  gap~$\Delta$ violates chirality conservation, and gives the resonance 
a finite width, $\Gamma\sim \Delta^2/\epsilon_c$; the resonance 
is quenched if the gap is larger than its energy~$\epsilon_c$. 
Thus, the resonance
can be controlled by a transverse field, and this can be
potentially employed to identify the signature of cloaked resonances
in the experimental data. 
The cloaked state can be also destroyed by magnetic field.
Indeed, the effect of the magnetic 
field~$B$ amounts to shifting the angular momentum. 
The resonance is destroyed when the magnetic flux through the cloaked
orbit becomes of the order of one flux quantum, or when the cyclotron 
energy is  comparable with the energy of the cloaked state. 
This analysis also shows that it may be desirable to avoid non-uniform
strain in the cloaked region 
to minimize the detrimental effects of pseudomagnetic 
fields~\cite{strain-field}. 

An interesting open question is the role of symmetry in this problem. 
Cloaked states have been found here for strong potentials with rotational
symmetry; earlier, cloaked states were identified for an arbitrarily weak 
potentials with  translational symmetry\cite{LL-Cloaking}.
The two solutions share a common feature: the two components of the 
wave function of the cloaked states are complex conjugate to 
each other: ~$\Psi = (\psi(x, y), \pm\psi^\ast(x, y))$. 
Indeed, if the Hamiltonian admits solutions of this type, one can 
show that the two chirality components are decoupled. Whether this can 
occur in other geometries, in particular, non-symmetric ones,
perhaps, with a different potential strength threshold, is not known 
to the author. 

Thus, we have shown that chirality of quasiparticles in bilayer graphene 
results in unconventional bound states that  
do not hybridize with the surrounding hole continuum. 
Such states occur if the confining potential exceeds the quantum
zero-point motion energy: $|U| > \hbar^2 / m^\ast a^2$. The cloaking 
can be implemented with both attractive and repulsive potentials. 
Trigonal  warping converts these states into resonances of finite
width, transverse electric and magnetic fields can be employed 
to quench these states. We predict that such states could be 
engineered in large clusters of charged impurities, and probed by 
scanning tunneling microscopy experiments. 
The author would like to thank L.S.Levitov, V.I.Falko, and P.Brouwer
for useful discussions. 
This research was supported by EPSRC/HEFCE No. EP/G036101.


\begin{thebibliography}{9}
\bibitem{MLG-dos-properties} A.~H.~Castro Neto et al,
Rev. Mod. Phys. {\bf 81}, 109 (2009).
\bibitem{Falko} E.~McCann, V.~I.~Falko, Phys.Rev.Lett. {\bf 96}, 086805
(2006).
\bibitem{Broken-Phases} E.~McCann and M.~Koshino, 
Rep.Progr.Phys. {\bf 76}, 056503 (2013), and references therein.
%R.~Nandkishore and L.~Levitov, Phys.Scr. {\bf T146}, 014011 (2012)
\bibitem{Klein-paradox} M.~I.~Katsnelson, K.~S.~Novoselov, A.~K.~Geim, 
Nature Physics {\bf 2}, 620 (2006).
\bibitem{LL-Cloaking} N.~Gu, M.~Rudner, and L.~S.~Levitov, 
Phys.~Rev.~Lett.{\bf 107}, 156603 (2011).
\bibitem{Coulomb-Atomic-Collapse} A.~V.~Shytov, M.~I.~Katsnelson,
L.~S.~Levitov, Phys.Rev.Lett.  {\bf 99},  246802 (2007).
\bibitem{Crommie-Supercritical} Y.~Wang et al, Science {\bf 340}, 734 (2013). 
\bibitem{Cloaking-Pendry} J.~B.~Pendry, D.~Schurig, and D.~R.~Smith, 
Science {\bf 312}, 1780 (2006).
\bibitem{Cloaking-Leonhardt} U.~Leonhardt,  Science {\bf 312}, 1777 (2006).
\bibitem{Campos-RBS} L.~C.~Campos et al, Nature Communications {\bf
3}, 1239 (2012). 
\bibitem{Peeters} A.~Matulis and F.~M.~Peeters, Phys. Rev. B {\bf 77},
115423 (2008).
\bibitem{LL-low-d} L.~D.~Landau and E.~M.~Lifshitz, {\it Quantum
Mechanics: Non-Relativistic Theory}, Sec.45, Vol. 3 (3rd ed),
Pergamon Press (1977). 
\bibitem{LL-bound} ibid., Sec.33.
\bibitem{hydrogen-2d} 
B.~Zaslow and M.~E.~Zandler, Am. J. Physics {\bf 35}, 1187 (1967).
\bibitem{Das-Sarma-Pi} E.~H.~Hwang and S.~Das Sarma, Phys.
Rev. Lett. {\bf 101}, 156802 (2008).
%X.L.Yang et al, Phys. Rev. A {\bf 43}, 1186 (1991)
\bibitem{Quantum-Corrals} M.~F.~Crommie, C.~P.~Lutz, and D.~M.~Eigler, 
Science {\bf 262}, 218 (1993).
\bibitem{Whisper-gallery} Y.~Zhao et al, Science {\bf 348}, 672 (2015). 
%E.J.Heller et al, Nature {\bf 369}, 464 (1994)
\bibitem{Gradstein} I.~S.~Gradstein and I.~M.~Ryzhik, ed. by A. Jeffrey, 
{\it Table of Integrals, Series, and Products (5th ed)}, Academic
Press, New York (1994), Eqs. 8.473 and 6.521.2.
\bibitem{Bilayer-Gap} Y.~Zhang et al, Nature {\bf 459}, 820 (2009)
\bibitem{strain-field} F.~Guinea, M.~I.~Katsnelson, and A.~K.~Geim, 
Nat. Phys. {\bf 6}, 30 (2010).
\end{thebibliography}
\end{document}